\documentclass[conference]{IEEEtran}

\usepackage{xcolor}
\usepackage{booktabs} 
\usepackage{graphicx} 
\usepackage{amsmath}
\usepackage{amsfonts}

\usepackage[utf8]{inputenc}
\usepackage[backend=biber, style=numeric, citestyle=numeric, maxbibnames=6, minbibnames=6]{biblatex} 
\usepackage{graphicx} 

\begin{document}

\title{Exploring Secure Machine Learning Through Payload Injection and FGSM Attacks on ResNet-50}

\author{
    \IEEEauthorblockN{
        Umesh Yadav\IEEEauthorrefmark{1},
        Suman Niroula\IEEEauthorrefmark{2},
        Gaurav Kumar Gupta\IEEEauthorrefmark{3}
        Bicky Yadav \IEEEauthorrefmark{4}
    }

\IEEEauthorblockA{
        \IEEEauthorrefmark{1}The University of Toledo\\
        Email: uyadav@rockets.utoledo.edu
    }
    \IEEEauthorblockA{
        \IEEEauthorrefmark{2}Youngstown State University\\
        Email: sniroula04@student.ysu.edu
    }
    \IEEEauthorblockA{
        \IEEEauthorrefmark{3}Youngstown State University \\
        Email: gkgupta@student.ysu.edu
    }
     \IEEEauthorblockA{
        \IEEEauthorrefmark{4} Islington College \\
        Email: yadavbicky99@gmail.com
    }

    }

\maketitle

\begin{abstract}
This paper investigates the resilience of a ResNet-50 image classification model under two prominent security threats: Fast Gradient Sign Method (FGSM) adversarial attacks and malicious payload injection. Initially, the model attains a 53.33\% accuracy on clean images. When subjected to FGSM perturbations, its overall accuracy remains unchanged; however, the model’s confidence in incorrect predictions notably increases. Concurrently, a payload injection scheme is successfully executed in 93.33\% of the tested samples, revealing how stealthy attacks can manipulate model predictions without degrading visual quality. These findings underscore the vulnerability of even high-performing neural networks and highlight the urgency of developing more robust defense mechanisms for security-critical applications.
\end{abstract}


\section{Introduction}

\IEEEPARstart{I}{n}  the modern landscape of cybersecurity, machine learning (ML) models, especially in areas like image classification, are increasingly integrated into systems where robustness and security are paramount. However, these models are highly susceptible to adversarial attacks, where small, crafted perturbations can lead to incorrect predictions and, in more severe cases, unauthorized access or manipulation of systems\cite{Biggio_2018}. Such attacks expose fundamental weaknesses in ML models, undermining their performance and reliability in critical applications.

Adversarial learning has attracted substantial attention in AI and ML research due to its potential to expose vulnerabilities in otherwise robust models. By adding minor, often imperceptible perturbations to input data, adversarial attacks can force models into making incorrect predictions\cite{Moosavi-Dezfooli_2017_CVPR}. This poses serious risks, especially in critical sectors such as autonomous driving and facial recognition. This report explores the impact of adversarial attacks on a pre-trained convolutional neural network (CNN) for image classification, specifically ResNet50. ResNet50 is a deep residual network with 50 layers, known for its high performance in image classification tasks\cite{he2015deepresiduallearningimage}.

Additionally, we will implement the Fast Gradient Sign Method (FGSM) attack to to demonstrate the susceptibility of even high-performing models to subtle perturbations. FGSM is a white-box attack that uses the gradient of the loss with respect to the input data to create adversarial examples\cite{goodfellow2015explainingharnessingadversarialexamples}.In this project, we explore the intersection of cybersecurity and adversarial attacks through the implementation of both payload injection and FGSM attacks on ResNet50. The goal is to assess how resilient such models are to attacks that combine adversarial noise with malicious payloads, highlighting the potential risks in real-world cybersecurity applications such as facial recognition, autonomous systems, and access control mechanisms.\\

\section{Motivation and Problem Defination}
As ML models continue to integrate into critical cybersecurity systems, the ability to exploit these models through adversarial techniques poses significant threats. A study predicts that by 2025, 30\% of cyberattacks will involve adversarial machine-learning tactics\cite{venturebeat2024}. Pre-trained models are susceptible to perturbations adversarial attacks, which can undermine trust in AI systems due to the lack of customized defensive measures during training. For example, in an autonomous driving scenario, a manipulated stop sign image could be misclassified, potentially leading to catastrophic consequences\cite{sitawarin2018dartsdeceivingautonomouscars}. In facial recognition systems, for example, a carefully crafted adversarial attack could allow an unauthorized user to bypass authentication systems, leading to severe security breaches\cite{sharif2016accessorize}\cite{9505665}.

Another issue is payload injection, where malicious inputs are crafted not only to fool the model but also to execute harmful actions, potentially compromising the entire system.  A study estimates that over 41\% of companies using AI report concerns about adversarial attacks on their systems \cite{venturebeat2024}. The combination of adversarial perturbations and malicious payloads exposes vulnerabilities in widely used ML models, especially those deployed in high-stakes environments like security access control, intrusion detection systems, and network defense mechanisms.

\subsection{Usages Examples and Scenarios}
The relevance of adversarial attacks is profound in domains where security and precision are critical. In facial recognition systems, adversarial attacks could enable unauthorized access. Similarly, adversarial noise in medical imaging could lead to incorrect diagnoses, potentially jeopardizing patient health. This project simulates adversarial conditions in a controlled setting, using a pre-trained image classification model, to better understand and address these vulnerabilities.
Below is the sequence followed in this project:

\begin{enumerate}
    \item \text{Start}
    \item \text{Load a clean image.}
    \item \text{Get a prediction on the clean image.}
    \item \text{Inject payload into the image.}
    \item \text{FGSM attack on both clean and injected images.}
    \item \text{Get a prediction on modified both images from step 5.}
    \item \text{Extract the payload from the injected image.}
\end{enumerate}

This workflow allows us to assess the model's performance under various attack scenarios and evaluate the effectiveness of payload injection and extraction

\subsection{Problem Definition}
The primary focus of this research is to evaluate the resilience of a pre-trained image classification model, ResNet50, to adversarial attacks that include both perturbations and malicious payload injection. We investigate how adversarial perturbations affect model accuracy and assess the success rate of payload injection attacks. Even minor perturbations can lead to significant misclassifications, and this project seeks to demonstrate how vulnerable popular pre-trained models like ResNet50 are in adversarial conditions.

\section{Related Work}
Research on adversarial attacks has been a focal point in machine learning (ML), particularly in the context of image classification and security-sensitive applications. The concept of adversarial examples was first introduced by Goodfellow et al. (2015), who demonstrated that even small, imperceptible perturbations to input data could lead to significant misclassifications in neural networks \cite{goodfellow2015explainingharnessingadversarialexamples}. This breakthrough spurred the development of various attack techniques, including the Fast Gradient Sign Method (FGSM), which uses the gradient of the loss function with respect to the input data to generate adversarial examples \cite{goodfellow2015explainingharnessingadversarialexamples}. Subsequent works such as the Projected Gradient Descent (PGD) method further refined adversarial attacks, making them more robust and harder to defend against \cite{madry2019deeplearningmodelsresistant}.

In cybersecurity, adversarial attacks have raised concerns about the security and robustness of machine learning models, especially when deployed in high-stakes environments. For example, in autonomous driving systems, adversarial attacks can manipulate stop sign images, causing vehicles to misinterpret critical road signs, as demonstrated by Sitawarin et al. (2018) \cite{sitawarin2018dartsdeceivingautonomouscars}. This illustrates how adversarial perturbations can lead to severe real-world consequences when they exploit vulnerabilities in image classification models. Similarly, adversarial examples have been shown to compromise facial recognition systems, allowing attackers to bypass security protocols and gain unauthorized access \cite{sharif2016accessorize}\cite{9505665}. The threat posed by adversarial examples extends beyond misclassifications, with the potential to enable malicious actions, such as manipulating security systems or causing critical failures in autonomous systems.

One of the significant advancements in adversarial attacks has been the integration of payload injection techniques. In this approach, adversarial examples not only mislead the model into making incorrect predictions but also embed malicious payloads that can compromise the entire system. For example, Yuan et al. (2019) explored how adversarial payloads can be injected into images, altering their underlying data without changing their visual appearance \cite{yuan2019adversarialexamplesdeep}. This allows attackers to manipulate systems without triggering alarms, making it a potent tool for bypassing security mechanisms. Furthermore, Shen et al. (2021) demonstrated the effectiveness of such attacks on deep learning-based face recognition systems, where adversarial stickers could be used to alter the model's predictions, leading to unauthorized access \cite{9505665}.

In addition to the attack techniques, the defense against adversarial attacks has been a significant area of research. Adversarial training, where models are trained on both clean and adversarial examples, has been a common defense strategy. However, many defense mechanisms are still vulnerable to adaptive attacks, which are specifically designed to circumvent these defenses \cite{athalye2018obfuscatedgradientsfalsesense}. For example, Madry et al. (2019) introduced adversarial training with Projected Gradient Descent (PGD) as a robust defense, although it has limitations against more sophisticated attacks \cite{madry2019deeplearningmodelsresistant}. Additionally, ensemble adversarial training, which combines adversarial examples from multiple sources, has been proposed as a defense mechanism, but it too has been shown to be vulnerable to evolving attack strategies \cite{tramer2017ensembleadversarialtraining}.

Adversarial attacks have also found significant applications in the healthcare domain. In medical imaging, adversarial perturbations can mislead diagnostic models, resulting in incorrect diagnoses that could jeopardize patient safety. Finlayson et al. (2019) demonstrated how adversarial attacks could affect medical machine learning models, making them a critical area of concern for healthcare applications \cite{finlayson2019adversarialattacksmedicalimaging}. These results underscore the need for robust and secure AI systems in critical sectors like healthcare, where the consequences of misclassifications could be life-threatening.

Further studies have explored the broader implications of adversarial attacks in security-critical applications such as autonomous vehicles, facial recognition, and access control systems. For example, the combination of adversarial perturbations and malicious payloads can enable attackers to alter the functionality of systems in subtle yet impactful ways. The ability to bypass authentication in facial recognition systems or manipulate the output of autonomous driving systems demonstrates the potential for adversarial attacks to cause significant harm to security infrastructures \cite{sharif2016accessorize}\cite{sitawarin2018dartsdeceivingautonomouscars}.

In response to these threats, researchers have focused on developing defensive mechanisms, including adversarial training and input sanitization. However, many of these defenses remain limited in their effectiveness, especially against adaptive attacks. For instance, Athalye et al. (2018) showed that models relying on obfuscated gradients could be easily bypassed with targeted attacks, emphasizing the need for more resilient defense strategies \cite{athalye2018obfuscatedgradientsfalsesense}. Despite the challenges in defending against adversarial attacks, progress has been made with techniques such as ensemble adversarial training and gradient masking, although further research is needed to create models that are truly robust against a wide range of attacks \cite{tramer2017ensembleadversarialtraining}.

In summary, while adversarial attacks have been extensively studied, the combination of adversarial perturbations and payload injection adds a new layer of complexity, making it even more challenging to defend against these attacks. The growing use of ML models in security-critical applications highlights the urgent need for more robust defense strategies that can handle both adversarial perturbations and malicious payloads simultaneously. Current defense strategies, although promising, are still vulnerable to evolving attack techniques, and future research must address these challenges to ensure the security and reliability of machine learning systems in high-stakes environments.

\section{Approach}
\subsection{Research Approach}
This project employs the Fast Gradient Sign Method (FGSM) to attack a pre-trained image classification model, ResNet50, trained on the ImageNet dataset. The steps include:

\begin{enumerate}
    \item \textbf{Model Selection:} Using a pre-trained ResNet50 model for image classification.
    
    \item \textbf{Baseline Evaluation:} Testing the model on clean, non-adversarial data to establish baseline accuracy.
    
    \item \textbf{Adversarial Example Generation:} FGSM is used to create adversarial examples by adding small perturbations in the direction of the loss gradient. The equation used is:
    \begin{equation}
        x_{\text{adv}} = x + \epsilon \cdot \text{sign}(\nabla_x J(\theta, x, y))
    \end{equation}
    where $x_{\text{adv}}$ is the adversarial image, $x$ is the original image, $\epsilon$ controls perturbation size, and $\nabla_x J$ is the gradient of the loss function.

    \item \textbf{Payload Injection:} Malicious payloads are injected into clean and the adversarially perturbed inputs. These payloads mimic potential cybersecurity attacks, such as modifying access controls or altering output predictions.
    
    \item \textbf{Evaluation of Adversarial and Payload Data:} The model is evaluated on data with adversarial perturbations and injected payloads, assessing both accuracy degradation and the success of the payload injection.
    
    \item \textbf{Comparison:} Performance on adversarial data is compared with baseline results to quantify the effect of the attack.
    
    \item \textbf{Payload Extraction:} The payload injected into the image is extracted to evaluate its effectiveness in compromising system security.
\end{enumerate}

\subsection{Justifications for the Approach}
FGSM was chosen due to its simplicity and effectiveness in generating adversarial examples. It allows for an illustration of how small perturbations can drastically affect a model's performance. By combining FGSM with payload injection, this approach demonstrates how adversarial attacks can not only misclassify inputs but also serve as a vector for compromising security systems. Using a pre-trained model like ResNet50 enables the focus to remain on evaluating adversarial resilience, without the overhead of training a new model from scratch.

\section{Evaluation}

\subsection{Research Questions and Objectives}
The primary objective of this evaluation is to assess the resilience of ResNet50, a pre-trained image classification model, against adversarial attacks, specifically the Fast Gradient Sign Method (FGSM), and to examine the effect of malicious payload injection on model predictions. The evaluation aims to answer the following research questions:

\begin{itemize}
    \item How does FGSM affect the confidence and accuracy of ResNet50 when exposed to adversarial perturbations?
    \item Does the model remain vulnerable when small, imperceptible adversarial perturbations are added?
    \item How successful is the payload injection technique in compromising the model’s predictions?
    \item Can the malicious payload be effectively extracted from the image, demonstrating its injection into the system?
\end{itemize}

\subsection{Evaluation Metrics}
The evaluation focuses on several key metrics:

\begin{itemize}
    \item \textbf{Confidence Levels}: The confidence of the model’s predictions on the original, payload-injected, and FGSM-attacked images is recorded. This allows us to compare how the model’s confidence changes before and after adversarial perturbations.
    
    \item \textbf{Prediction Class Changes}: By observing the model’s predicted class across different stages (original, FGSM, payload), we can determine how vulnerable the model is to being fooled.
    
    \item \textbf{Success Rate of Payload Injection/Extraction}: For each image, the success of embedding a payload and later extracting it is measured, ensuring that the injection did not disrupt the image’s visual integrity. The success rate is calculated as:
    
    \begin{equation}
        \text{Success Rate} = \frac{\text{Number of Successful Extractions}}{\text{Total Number of Images}} \times 100
    \end{equation}

    \item \textbf{Accuracy Before and After FGSM}: By comparing the model’s accuracy before and after FGSM attacks, we can quantify the impact of adversarial perturbations on the model’s ability to correctly classify images. The baseline accuracy is calculated as:






\begin{equation}
\small
\text{Baseline Acc.} = \frac{\text{Correct}}{\text{Total}} \times 100
\end{equation}

\begin{equation}
\small
\text{Post-FGSM Acc.} = \frac{\text{Correct}_{\text{FGSM}}}{\text{Total}} \times 100
\end{equation}

\begin{equation}
\small
\Delta \text{Acc.} = \text{Baseline Acc.} - \text{Post-FGSM Acc.}
\end{equation}

\end{itemize}

\section{Results}

The evaluation of ResNet50's resilience against adversarial attacks and payload injection was conducted using a set of 15 images, with adversarial examples generated using the Fast Gradient Sign Method (FGSM). The following key results were observed:

\subsection{Original Accuracy}
The ResNet50 model exhibited a baseline accuracy of \textbf{53.33\%} on clean, non-adversarial images. This was determined by comparing the model’s predicted class with the actual class (manually verified) for each image. This accuracy serves as a benchmark for evaluating the impact of adversarial attacks and payload injections.

\subsection{Impact of FGSM Attack}
After applying the FGSM attack on the original images, the model's accuracy remained unchanged at \textbf{53.33\%}. However, one of the most significant findings was the increase in confidence levels observed after the FGSM attack. In \textbf{100\%} of the cases, the model became more confident in its predictions, even when the predictions were incorrect.

This increase in confidence can be attributed to the nature of targeted adversarial attacks. FGSM specifically creates perturbations by adjusting the pixel values in the direction of the model’s loss gradient. This forces the model to move towards a specific, incorrect classification. The adversarial perturbations exploit the underlying weaknesses in the model’s decision boundaries, which not only causes misclassifications but also inflates the model’s confidence in those wrong predictions\cite{goodfellow2015explainingharnessingadversarialexamples}\cite{kurakin2017adversarialexamplesphysicalworld}.

In essence, FGSM steers the model towards a different region in its decision space, where it "believes" the adversarially perturbed input belongs to a particular class. Since the attack is crafted to maximize the model’s loss with respect to a chosen class, it naturally leads to higher confidence scores. The model interprets the perturbed input as being more representative of the incorrect class, resulting in a misleading increase in confidence\cite{szegedy2014intriguingpropertiesofneuralnetworks}\cite{moosavidezfooli2017universaladversarialperturbations}.

\subsection{Impact of FGSM Attack on Payload-Injected Images}
When the FGSM attack was applied to the payload-injected images, the model's accuracy also remained at \textbf{53.33\%}. However, as with the original images, \textbf{93.33\%} of the cases exhibited an increase in confidence levels post-attack. This further supports the notion that FGSM, as a targeted attack, not only disrupts the model's ability to correctly classify the image but also artificially boosts the model’s confidence in its incorrect predictions. The combination of payload injection and FGSM demonstrates the potential risks when adversarial techniques are used to manipulate both the model’s prediction and its certainty.

\subsection{Payload Injection Success}
The success rate of the payload injection was \textbf{93.33\%}, meaning that in the majority of cases, the payload was successfully injected into the image and later extracted without visibly altering the image or affecting the model's prediction.

\subsection{Confidence Levels and FGSM}
The most notable trend observed during the evaluation was the consistent increase in confidence levels after applying the FGSM attack. In both the original images and payload-injected images, the model became significantly more confident in its predictions, even when they were incorrect. This behavior is a hallmark of adversarial attacks like FGSM, where the perturbations exploit the model’s learned patterns and decision boundaries. The targeted nature of FGSM forces the model to treat the manipulated input as being more representative of a class, which artificially inflates its confidence levels.

This trend is particularly concerning in real-world scenarios where adversarial attacks can be used to manipulate machine learning systems into confidently making wrong predictions. The combination of higher confidence in incorrect classifications could have serious implications, especially in security-critical applications such as facial recognition or autonomous driving.

\subsection{Summary of Results}
\begin{itemize}
    \item Baseline Accuracy: \textbf{53.33\%}
    \item Post-FGSM Accuracy: \textbf{53.33\%}
    \item Post-FGSM Accuracy on Payload-Injected Images: \textbf{53.33\%}
    \item Confidence Increase after FGSM (Original Images): \textbf{100\%}
    \item Confidence Increase after FGSM (Payload-Injected Images): \textbf{93.33\%}
    \item Payload Injection Success Rate: \textbf{93.33\%}
\end{itemize}

\begin{figure}
    \centering
    \includegraphics[width=0.95\linewidth]{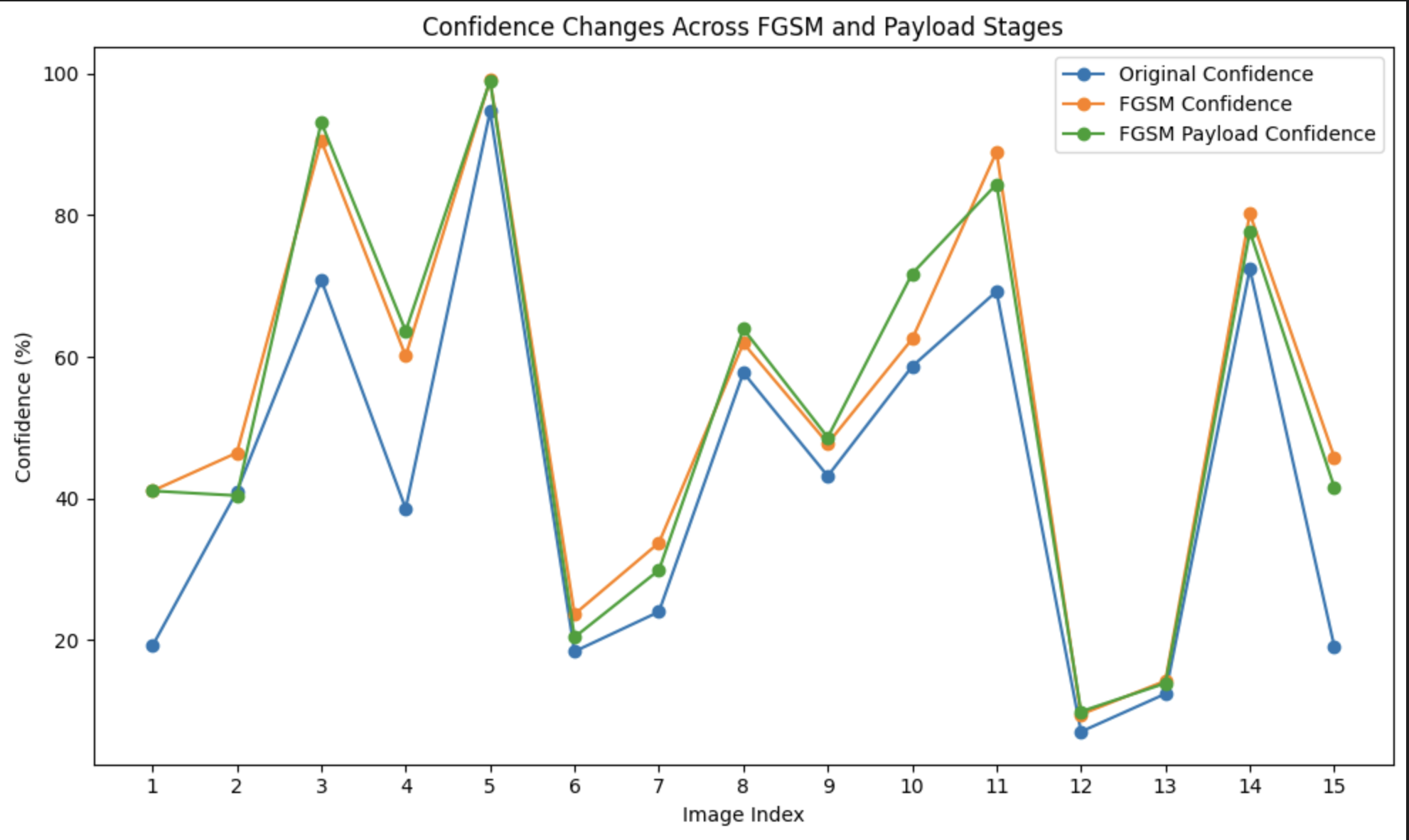}
    \caption{Comparison Of Confidence Levels Before And After FGSM attack}
    \label{fig:enter-label}
\end{figure}

Overall, while the FGSM attack did not significantly reduce the model’s classification accuracy, it led to an inflated confidence in incorrect predictions. The targeted nature of FGSM means that the perturbations were specifically designed to increase the model's loss in favor of a particular class, which often resulted in an exaggerated confidence in those wrong predictions. The payload injection technique was also largely successful, with most payloads being successfully injected and extracted without visibly altering the image.

\section{Discussion and Future Work}

\subsection{Discussion}
The results from this study highlight key insights into the behavior of ResNet50 when exposed to targeted adversarial attacks (FGSM) and payload injection techniques. One of the most significant findings is the increase in confidence levels after the FGSM attack. This behavior demonstrates that while the model's overall accuracy did not drop significantly, the adversarial perturbations caused the model to become more confident in its misclassifications. This is a concerning trend, particularly in real-world applications where models may be making high-stakes decisions based on their confidence in predictions.

The nature of FGSM, which adjusts pixel values in the direction of the model's loss gradient, forces the model into a different region of its decision space. This leads the model to perceive the adversarially perturbed input as strongly resembling the target class, which inflates its confidence, even when the classification is incorrect \cite{goodfellow2015explainingharnessingadversarialexamples} \cite{szegedy2014intriguingpropertiesofneuralnetworks}. This effect was observed in both original and payload-injected images, underscoring the vulnerability of deep learning models to targeted adversarial attacks.

Moreover, the payload injection technique proved largely successful, with a \textbf{93.33\%} success rate in extracting the payload from the images. This indicates that adversarial attacks combined with payload manipulation can be used to subtly alter the functionality of models without alerting users to the manipulation. 

\subsection{Limitations}
One limitation of this study is that the FGSM attack did not significantly impact the model’s accuracy, which remained at \textbf{53.33\%} across all tests. While the increase in confidence levels highlights a critical weakness, future work could explore stronger adversarial techniques, such as Projected Gradient Descent (PGD) or Carlini \& Wagner (C\&W) attacks \cite{madry2019deeplearningmodelsresistant} \cite{carlini2017towards}. These methods could potentially have a greater impact on both accuracy and confidence, providing a more comprehensive understanding of the model’s vulnerabilities.

Another limitation is the relatively small dataset of 15 images, which may not fully capture the model’s performance across a wide range of scenarios. A larger dataset with more diverse images could provide more generalized insights into the model's susceptibility to adversarial attacks.

\subsection{Future Work}
Future research could focus on exploring different adversarial attack techniques, such as PGD or iterative FGSM, to measure the model’s robustness under more sophisticated perturbations. Additionally, applying these attacks to other pre-trained models, such as Inception or VGGNet, could reveal whether certain architectures are more resilient to adversarial attacks.

Moreover, integrating defensive mechanisms, such as adversarial training or input sanitization, could be explored to mitigate the effects of adversarial perturbations. These methods could be tested for their effectiveness in reducing both confidence inflation and misclassification rates.

Finally, expanding the scope of this research to include real-time applications, such as autonomous vehicles or facial recognition systems, could provide valuable insights into how adversarial attacks might affect high-stakes systems in practice.

\nocite{*}
\printbibliography
\end{document}